\documentclass[prb,reprint,twocolumn,showpacs]{revtex4}
\usepackage{amsmath,amssymb,graphicx,xcolor,units,hyperref}

\newcommand{\dd}[1]{\mathrm{d}#1\,}

\newcommand{\bra}[1]{\langle{#1}\rvert}
\newcommand{\ket}[1]{\lvert{#1}\rangle}

\renewcommand{\vec}[1]{\mathbf{#1}}

\newcommand\varpm{\mathbin{\vcenter{\hbox{%
  \oalign{\hfil$\scriptstyle+$\hfil\cr
          \noalign{\kern-.3ex}
          $\scriptscriptstyle({-})$\cr}%
}}}}
\newcommand\varmp{\mathbin{\vcenter{\hbox{%
  \oalign{$\scriptstyle({+})$\cr
          \noalign{\kern-.3ex}
          \hfil$\scriptscriptstyle-$\hfil\cr}%
}}}}

\begin{document}

\title{Energy transport via multiphonon processes in graphene}

\author{P. Virtanen}
\affiliation{O.V. Lounasmaa Laboratory, Aalto University,
  P.O. Box 15100,FI-00076 AALTO, Finland}

\date{\today}


\begin{abstract}
  The Dirac dispersion of graphene limits the phase space available
  for energy transport between electrons and acoustic phonons at
  temperatures above the Bloch-Gr\"uneisen temperature.  Consequently,
  energy transport can be dominated by supercollision events,
  involving also other scattering processes.  Scattering from flexural
  phonons can compensate for the large momentum transfer involved in
  scattering from thermal acoustic phonons, and enables similar
  supercollision events as disorder. Such multiphonon processes are
  also allowed by selection rules. I show that acoustic-flexural
  process can in the energy transport be of the same order of
  magnitude as direct flexural and acoustic phonon processes,
  depending on electronic screening and mechanical strain.
\end{abstract}

\maketitle

\section{Introduction}

Electron-phonon interaction in graphene is characterized on the one
hand by its relative weakness as compared to other materials, but on
the other also by the richness of different possible processes playing
a role in it.  Understanding the electron-phonon interaction and its
ability to transfer energy in graphene is moreover of fundamental
importance for certain device applications, for instance in radiation
detection in which controlling the transport of energy is important.

Recent theoretical and experimental results have highlighted that
phase space constraints and their breaking
\cite{song2012-dae,betz2013-sci,chen2012-epm} have a large importance
for energy transfer at intermediate temperatures. Above the
Bloch-Gr\"uneisen temperature $T_{BG} = \hbar k_F s/k_B$, which in
graphene can be fewer than tens of Kelvins, the wave vector
$q_T=k_BT/\hbar s$ of thermal phonons is larger than the electronic
Fermi wave vector $k_F$ defined within a single Dirac cone.  As the
sound velocity $s\approx2\times\unit[10^4]{m/s}$ in graphene is small
compared to the Fermi velocity $v_F\approx\unit[10^6]{m/s}$, only a
small fraction of phonon and electron states inside the thermal window
$|\hbar v_F|\vec{k}| -\mu|, \hbar s |\vec{q}| \lesssim{}T$ around
Fermi level $\mu$ satisfy the constraint $\vec{k}'-\vec{k}=\vec{q}$
required for momentum-conserving scattering of an electron from
$\vec{k}$ to $\vec{k}'$. This suppresses the energy flow via direct
electron-phonon interaction, allowing processes mediated by disorder
or other additional scattering processes to dominate,
\cite{song2012-dae,song2011-sab} up to and even above temperatures
where optical phonons activate.

In ultraclean suspended graphene samples it is possible to achieve
mean free paths long enough to see the effect of electron-phonon
scattering on the electrical resistance.
\cite{bolotin2008-tdt,castro2010-loc} The disorder-mediated energy
transfer mechanism weakens as the scattering length increases, and
intrinsic processes can start to compete with it. For resistance, an
important process turns out to be scattering from flexural
phonons. \cite{mariani2008-fpi,castro2010-loc,song2012-dae} 

\begin{figure}
  \includegraphics{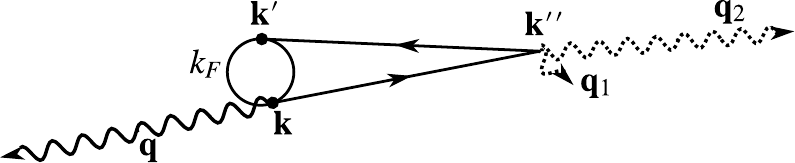}
  \caption{\label{fig:scattering} Multiphonon process in which
    electron scatters from $\vec{k}$ to $\vec{k}'$ on the Fermi circle $k_F$
    via an intermediate state. It first emits a large-momentum acoustic
    phonon $\vec{q}$ and
    scatters back to Fermi surface from flexural phonons
    $\vec{q}_1$, $\vec{q}_2$.  The energy transferred to the phonons
    in the process is approximately that of the acoustic phonon,
    $\omega_{\vec{q}}$, since the flexural phonon frequencies on the relevant
    wave vector scale are small.
  }
\end{figure}

In this work, I discuss how the phonon processes visible in the
resistance contribute to supercollision processes in energy transport.
Only multiphonon processes that couple to electrons via gauge
potential \cite{guinea2008-gfi,morozov2006-ssw} are in general allowed
by effective selection rules. \cite{song2011-sab} An important process
turns out to be the combined process in which a large-energy acoustic
phonon is emitted, and momentum balance is obtained via quasielastic
scattering from low-energy flexural phonons [see
Fig.~\ref{fig:scattering}]. This is equivalent to supercollisions
involving scattering from dynamical ripples in the graphene. In the
degenerate regime, this results to a $G_{\rm
  e-ph}\propto{}T^2\ln^2(T)$ temperature dependence
[Eq.~\eqref{eq:flex-ac}] in the electron-phonon thermal
conductivity. For typical parameters of ultraclean suspended graphene
samples at charge densities around $n\approx\unit[10^{11}]{cm^{-2}}$,
the process can compete with the energy flow transferred via direct
flexural or acoustic phonon coupling [see Fig.~\ref{fig:GJ}].  The
effective scattering length can be related to the additional
resistance from flexural phonon scattering
[Eq.~\eqref{eq:resistance-tau}], which enables a cross-check on the
mechanism for the dissipated power density.  Gauge potential coupling
also allows supercollision events involving longitudinal and
transverse acoustic phonons. This process turns out to be less
important than that involving flexular phonons.

\section{Model}

The monolayer graphene Hamiltonian under consideration is
\begin{align}
  H &= H_e + H_{ph} + H_{e-ph}
  \,,
  \\
  H_e &= 
  \sum_{\vec{k}} 
  \Psi^\dagger_{\vec{k}}(\hbar v_F \vec{k}\cdot\vec{\sigma})\Psi_{\vec{k}}
  \,,
\end{align}
which includes phonons and their coupling to electrons:
\begin{align}
  H_{ph}
  &=
  \sum_{\vec{q}; j=L,T,LO,TO,F} \hbar\omega_{j,\vec{q}}b_{j,\vec{q}}^\dagger b_{j,\vec{q}}
  \,,
  \\
  H_{e-ph}
  &=
  \sum_{\vec{q}\vec{k};j=L,T,LO,TO}
  \Psi_{\vec{k}+\vec{q}}^\dagger
  M_{j,\vec{q}}
  \Psi_{\vec{k}}
  (b_{j,\vec{q}} + b_{j,-\vec{q}}^\dagger)
  \\\notag
  &
  +
  \sum_{\vec{q}_1\vec{q}_2\vec{k}}
  \Psi_{\vec{k}+\vec{q}_1+\vec{q}_2}^\dagger
  M_{F,\vec{q}_1\vec{q}_2}
  \Psi_{\vec{k}}
  \\\notag
  &\qquad\times
  (b_{F,\vec{q}_1} + b_{F,-\vec{q}_1}^\dagger)
  (b_{F,\vec{q}_2} + b_{F,-\vec{q}_2}^\dagger)
  \,.
\end{align}
Here, $\Psi_{\vec{k}}=(a_{A,\vec{k}}, a_{B,\vec{k}})^T$ is the electron
pseudospinor operator in the sublattice $A/B$ space.  We consider here
only spin-independent processes within a single valley of the
graphene, so that observables need to be multiplied with the corresponding
multiplicity $N=4$ of independent electron species.

The spectrum of the longitudinal (L) and transverse (T) acoustic modes
are taken as $\omega_{L/T,\vec{q}}=s_{L/T}|\vec{q}|$, where
$s_L=\unit[2.1\times{}10^4]{m/s}$, $s_T=\unit[1.4\times{}10^4]{m/s}$
are the sound velocities. Optical phonons (LO, TO) have
$\omega_{LO/TO}\approx\unit[200]{meV}$.  The flexural (F) phonon
spectrum in unstrained graphene is
$\omega_{F,\vec{q}}\approx\alpha{}q^2[1 + (q_c/q)^2]^{\eta/4}$, with
$\alpha=\unit[4.6\times{}10^{-7}]{m^2/s}$ and $\eta\approx1$ (the
precise value of the exponent affects prefactors in the results
below). The infrared cutoff
$q_c\approx\unit[0.1]{\AA^{-1}}\sqrt{T/\unit[300]{K}}$ arises from
anharmonic effects. \cite{mariani2008-fpi,zakharchenko2010-scs} The
flexural phonon energies are smaller than those of the acoustic modes
up to a high temperature $\hbar s_L^2/k_B\alpha \approx
\unit[7300]{K}$.

The electron-phonon coupling matrixes $M$ have been discussed previously e.g. in
Refs.~\onlinecite{mariani2008-fpi,bistritzer2009-eci,ochoa2011-tdr}, and we use the same
models here:
\begin{subequations}
\label{eq:monolayer-e-ph}
\begin{align}
  M_{L,\vec{q}}
  &=
  i|\vec{q}|
  \sqrt{
    \frac{\hbar}{2{\cal V}\rho\omega_{L,\vec{q}}}
  }
  \begin{pmatrix}
    D_1(|\vec{q}|)
    &
    -iD_2 e^{2i\theta_\vec{q}}
    \\
    i D_2 e^{-2i\theta_\vec{q}}
    &
    D_1(|\vec{q}|)
  \end{pmatrix}
  \,,
  \\
  M_{T,\vec{q}}
  &=
  i|\vec{q}|
  \sqrt{
    \frac{\hbar}{2{\cal V}\rho\omega_{T,\vec{q}}}
  }
  \begin{pmatrix}
    0
    &
    D_2 e^{2i\theta_\vec{q}}
    \\
    D_2 e^{-2i\theta_\vec{q}}
    &
    0
  \end{pmatrix}
  \,,
  \\
  M_{F,\vec{q}\vec{q}'}
  &=
  \frac{
    -
    \hbar
    |\vec{q}|
    |\vec{q}'|
  }{
    4{\cal V}\rho
    \sqrt{
      \omega_{F,\vec{q}}
      \omega_{F,\vec{q}'}
    }
  }
  \\\notag
  &\;\times
  \begin{pmatrix}
    D_1(|\vec{q} + \vec{q}'|)
    \cos(\theta_{\vec{q}} - \theta_{\vec{q}'})
    &
    -iD_2
    e^{i(\theta_{\vec{q}} + \theta_{\vec{q}'})}
    \\
    c.c.
    &
    c.c.
  \end{pmatrix}
  \,,
  \\
  M_{opt,\vec{q}}
  &=
  \frac{2D_2}{a}
  \sqrt{\frac{\hbar}{2 {\cal V}\rho \omega_{\rm opt}}}
  \begin{pmatrix}
    0
    &
    -f_{LO/TO}
    e^{-i\theta_{\vec{q}}}
    \\
    -\mathrm{c.c.}
    &
    0
  \end{pmatrix}
  \,.
\end{align}
\end{subequations}
Here, ${\cal V}$ is the area of the graphene sheet, and $\rho$ its
mass density.  $D_1(q)=D_1/\epsilon_r(q)$ is the screened deformation
potential, and $D_2=\hbar{}v_F\beta/(2a)$ the gauge potential, where
$\beta\approx2\ldots3$ is a dimensionless parameter, and $a$ the
equilibrium distance between carbon atoms.  The factor $f_{LO}=1$ for
LO phonons and $f_{TO}=i$ for TO phonons.  The angles are
$\theta_{\vec{q}}=\arctan(q_y/q_x)$.  We take electronic screening of
the deformation potential into account within a static
Thomas-Fermi type approximation in the matrix element,
$\epsilon_r(q)=1 +
\frac{e^2}{\pi\hbar{}v_F\epsilon_0}\frac{k_F}{q}$. \cite{castro2010-loc}
Note that while the $D_1$ coupling is an identity matrix in the
sublattice space, this is not the case for the $D_2$ gauge coupling.

\section{Multiphonon processes}

We now consider the scattering rate for a similar supercollision
process as discussed in Ref.~\onlinecite{song2012-dae}: an electron
from a state $\vec{k}$ close to the Fermi level emits an acoustic
phonon, ends up in a virtual state at a large momentum
$\vec{k}''=\vec{k}-\vec{q}\approx{}-\vec{q}$,
$|\vec{q}|\sim{}k_BT/\hbar s_L\gg{}k_F$, and scatters back to a state
$\vec{k}'$ at the Fermi level by interacting with other phonons (see
Fig.~\ref{fig:scattering}).

The rate of such events is found via a standard T-matrix calculation,
\cite{bruus2004-mbq}
$W_{fi}=\frac{2\pi}{\hbar}|\bra{f}T(\epsilon_i)\ket{i}|^2\delta(\epsilon_f-\epsilon_i)$. Expanding
in $H_{\rm e-ph}$, we have $T(\epsilon) = H_{\rm e-ph} + H_{\rm
  e-ph}[\epsilon + i0^+ - H_{e}-H_{ph}]^{-1}H_{\rm e-ph} +
\ldots=T_1+T_2+\ldots$. Straightforward calculation along similar lines
as discussed in Refs.~\onlinecite{song2012-dae,song2011-sab} gives the
second-order matrix element for electron scattering from $\vec{k}$ to
$\vec{k}'\ne\vec{k}$:
\begin{align}
  \label{eq:element}
  \bra{f}T_2(E_i)\ket{i}
  &=
  \frac{1}{2}
  \sum_{\alpha\vec{k}\alpha'\vec{k}'}
  \sum_{qq'}
  w^{qq'}_{\alpha'\vec{k}'\alpha\vec{k}}
  \delta_{\vec{k}'-\vec{k},\vec{Q}_q+\vec{Q}_{q'}}
  \\\notag
  &\quad\times
  \bra{f} u_{q'} u_q\ket{i} (1-n_{\alpha'\vec{k}'}) n_{\alpha\vec{k}} 
  \\
  \label{eq:rate}
  w^{qq'}_{\alpha'\vec{k}',\alpha\vec{k}}
  &\simeq
  \Phi_{\alpha'\vec{k'}}^\dagger
  \Bigl(
  M_{q'}
  \frac{1}{
    \hbar v_F \vec{Q}_q\cdot\sigma
  }
  M_{q}
  \\\notag
  &\qquad+
  M_{q}
  \frac{1}{
    -\hbar v_F\vec{Q}_q\cdot\sigma
  }
  M_{q'}
  \Bigr)
  \Phi_{\alpha\vec{k}}
  \,.
\end{align}
Here $n$ are occupations of electron eigenstates
$(\alpha=\pm1,\vec{k})$ in the initial state, which correspond to the
single-particle pseudospinors $\Phi_{\alpha\vec{k}}=2^{-1/2}(\alpha
e^{-i\theta_{\vec{k}}}, 1)^T$.  Also, $q$ labels phonon degrees of
freedom: for acoustic phonons $q=(L/T, \vec{q})$, $\vec{Q}_q=\vec{q}$,
$u_q=b_{j\vec{q}}+b_{j,-\vec{q}}^\dagger$, and for flexural phonons
$q=(\vec{q}_1, \vec{q}_2)$, $\vec{Q}_q=\vec{q}_1+ \vec{q}_2$,
$u_q=(b_{F\vec{q}_1}+b_{F,-\vec{q}_1}^\dagger)(b_{F\vec{q}_2}+b_{F,-\vec{q}_2}^\dagger)$.
As in Ref.~\onlinecite{song2012-dae}, we assumed the intermediate
state $\vec{k}''=\vec{k}+\vec{Q}_q\simeq{}\vec{Q}_q$ lies at a high
energy so that we can neglect the other terms in the denominator.
This assumption is valid for $|\vec{k}''|\gg{}k_F$, and relies on the
energy scale separation between the electrons and the phonons.  Due to
this approximation, the matrix element is also the same both for
phonon emission and absorption processes.

The sign change in the second term of Eq.~\eqref{eq:rate} is due to
the electronic structure of the monolayer graphene and momentum
conservation. As observed in Ref.~\onlinecite{song2011-sab}, this
results to an effective selection rule that prevents many multiphonon
processes. However, based on Eq.~\eqref{eq:monolayer-e-ph}, we can see
that certain processes involving gauge potential and having
distinguishable final states are still allowed.

Substituting in the matrix elements from Eq.~\eqref{eq:monolayer-e-ph}
and computing the averages over the initial and final state angles
$\theta_{\vec{k}}, \theta_{\vec{k}'}$ we find that matrix elements
such as $w^{L,L}$ and $w^{T,T}$, $w^{L,TO}$ vanish. The angle-averaged
matrix elements that remain are:
\begin{align}
  |\bar{w}^{L,F}|^2
  &=
  \frac{\hbar D_{1}(|\vec{q}|)^2D_2^2}{4{\cal V}^2 v_F^2 \rho^3 s_L|\vec{q}|}
  \Bigl[
  \cos^2(3\theta_{\vec{q}})\cos^2(\theta_{\vec{q}_1}-\theta_{\vec{q}_2})
  \\\notag
  &\quad
  +\cos^2(\theta_{\vec{q}}+\theta_{\vec{q}_1}+\theta_{\vec{q}_2})
  \Bigr]
  \frac{
    |\vec{q}_1|^2|\vec{q}_2|^2
  }{
     \omega_{F,\vec{q}_1} \omega_{F,\vec{q}_2}
  }
  \,,
  \\
  |\bar{w}^{T,F}|^2
  &=
  \frac{
    \hbar
    D_{1}(|\vec{q}|)^2D_2^2
    |\vec{q}_1|^2|\vec{q}_2|^2
    \cos^2(3\theta_{\vec{q}})\cos^2(\theta_{\vec{q}_1}-\theta_{\vec{q}_2})
  }{4{\cal V}^2 v_F^2 \rho^3 \omega_{F,\vec{q}_1} \omega_{F,\vec{q}_2} s_T|\vec{q}|}
  \\
  |\bar{w}^{L,T}|^2
  &=
  \frac{1}{2{\cal V}^2 v_F^2 \rho^2 s_L s_T}
  D_1^2
  D_2^2
  \cos^2(3\theta_{\vec{q}})
  \,,
  \\
  |\bar{w}^{L,LO}|^2
  &=
  \frac{2}{{\cal V}^2\rho^2 v_F^2 s_L \omega_{LO}}
  \frac{1}{|\vec{q}|}
  \frac{1}{a^2}
  D_1^2 D_{2}^2
\end{align}
They exhibit the 6-fold rotation symmetry of the graphene lattice. The
angle factors $\cos^2(3\theta_{\vec{q}})$ and
$\cos^2(\theta_{\vec{q}}+\theta_{\vec{q}_1}+\theta_{\vec{q}_2})$
average to $1/2$ under global rotations.

The total power density carried by a processes involving flexural
phonons and the acoustic mode $j=L/T$ is
\begin{align}
  {\cal J}_{j}
  &=
  2\pi N
  \sum_{\gamma\gamma_1\gamma_2=\pm}
  \int_{-\infty}^\infty\dd{\xi}\dd{\xi'}\nu_1(\xi)\nu_1(\xi')
  \int\frac{\dd{^2q}\dd{^2q_1}}{(2\pi)^4}
  \\\notag
  &\;\times
  {\cal V}^2
  |\bar{w}^{j,F}|^2
  \delta(\xi'-\xi+\gamma\omega_{j,\vec{q}}+\gamma_1\omega_{F,\vec{q}_1}+\gamma_2\omega_{F,\vec{q}_2})
  \\\notag
  &\;\times
  (N_{\omega_{j,\vec{q}}} + \delta_{\gamma,+})
  (N_{\omega_{F,\vec{q}_1}} + \delta_{\gamma_1,+})
  (N_{\omega_{F,\vec{q}_2}} + \delta_{\gamma_2,+})
  \\\notag
  &\;\times
  (\gamma\omega_{j,\vec{q}}+\gamma_1\omega_{F,\vec{q}_1}+\gamma_2\omega_{F,\vec{q}_2})
  (1-n_{\xi'})n_{\xi}\rvert_{\vec{q}_2=-\vec{q}-\vec{q}_1}
  \,.
\end{align}
Here, $N_\omega=[e^{\hbar\omega/k_BT}-1]^{-1}$ are Bose functions, and
$n_\xi=[e^{(\xi-\mu)/k_BT_e}+1]^{-1}$ Fermi distributions of the
electrons in the graphene.  $\nu_1(E)=|E|/(2\pi\hbar^2v_F^2)$ is the
density of electron states per valley per spin. $\gamma=\pm$ denote
phonon emission and absorption. In the degenerate
limit, $|\mu|\gg{}T$, we have
\begin{align}
  \label{eq:Jdegenerate}
  {\cal J}_{j}
  &=
  2\pi N
  \nu_1(\mu)^2
  \sum_{\gamma\gamma_1\gamma_2=\pm}
  \int\frac{\dd{^2q}\dd{^2q_1}}{(2\pi)^4}
  \\\notag
  &\;\times
  {\cal V}^2
  |\bar{w}^{j,F}|^2
  (\gamma\omega_{j,\vec{q}}+\gamma_1\omega_{F,\vec{q}_1}+\gamma_2\omega_{F,\vec{q}_2})^2
  \\\notag
  &\;\times
  (N_{\omega_{j,\vec{q}}} + \delta_{\gamma,+})
  (N_{\omega_{F,\vec{q}_1}} + \delta_{\gamma_1,+})
  (N_{\omega_{F,\vec{q}_2}} + \delta_{\gamma_2,+})
  \\\notag
  &\;\times
  N^{el}_{\gamma\omega_{j,\vec{q}}+\gamma_1\omega_{F,\vec{q}_1}+\gamma_2\omega_{F,\vec{q}_2}}
  \rvert_{\vec{q}_2=-\vec{q}-\vec{q}_1}
  \,,
\end{align}
where $N^{el}_\omega=[e^{\omega/T_e}-1]^{-1}$.

We now make use of the energy scale separation between acoustic and
flexural phonons, to approximate
$\omega_{L/T}\pm\omega_F\approx\omega_{L/T}$. The main contribution to
the integral comes from wave vector combinations where
$\hbar\omega_F\ll{}k_B{}T$, so we also approximate
$N_{\omega_{F}}+1\approx{}N_{\omega_{F}}$.  Scattering from flexural
phonons is approximately elastic. The relevant integral over $\vec{q}_1$ is
\begin{align}
  \label{eq:Ycorr}
  {\cal Y}(\vec{q})
  &=
  \int\frac{\dd{^2q_1}}{(2\pi)^2}
  \frac{
    |\vec{q}_1|^2|\vec{q}_2|^2
    N_{\omega_{F,\vec{q}_1}}
    N_{\omega_{F,\vec{q}_2}}
  }{
    \omega_{F,\vec{q}_1} \omega_{F,\vec{q}_2}
  }
  \rvert_{\vec{q}_2=-\vec{q}-\vec{q}_1}
  \\\notag
  &\simeq
  \frac{
    q_{F,T}^4
  }{
    \alpha^2
    \pi
  }
  \frac{
    \log\bigl(
    \frac{|\vec{q}|}{q_c}
    +
    \frac{q_c}{|\vec{q}|}
    \bigr)
    +
    1.3
  }{
    |\vec{q}|^2 + 2q_c^2
  }
  \,,
\end{align}
where $q_{F,T}=\sqrt{\frac{k_BT}{\hbar\alpha}}$ is the wave vector of
thermal flexural phonons. The approximate result is a composite of the
asymptotic behavior for $q_c,|\vec{q}|\ll{}q_{F,T}$ and
$q_c\ll|\vec{q}|$ or $q_c\gg{}|\vec{q}|$. Including an angle factor
$\cos^2(\theta_{\vec{q}_1}-\theta_{-\vec{q}-\vec{q}_1})$ in the
integral gives the same result multiplied by $\approx{}0.5$, so that
the prefactors for the L and T cases are $c_L\approx1.5$,
$c_T\approx0.5$. If flexural phonons are at a temperature close to
that of the electrons, $q_{F,T}\gg{}k_F,q_{L/T,T}$, and the above
approximation can be used in the whole range.

\begin{figure}
  \includegraphics{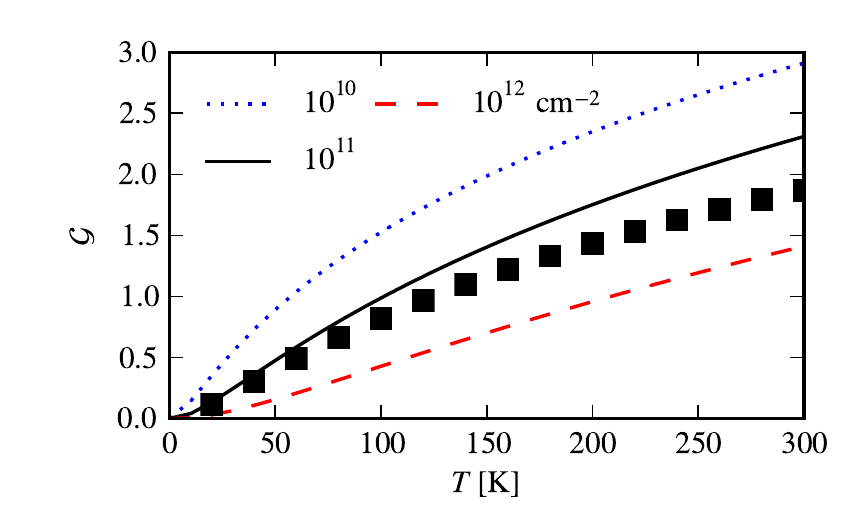}
  \caption{\label{fig:Gfunc}
    The dimensionless factor ${\cal G}$ in Eq.~\eqref{eq:flex-ac} as a function of 
    charge density and temperature. 
    Squares indicate results from Eq.~\eqref{eq:Jdegenerate} without further
    approximations (for $j=L$ and $n=\unit[10^{11}]{cm^{-2}}$).
  }
\end{figure}

The above results to the total power density
\begin{align}
  \label{eq:flex-ac}
  {\cal J}_{j}
  &\simeq
  N 
  \frac{
    D_1^2
    \nu_1(\mu)^2
  }{
    \rho^2 \alpha^3
  }
  k_B^3
  T^2
  \Delta T
  \frac{
    c_j
    D_2^2
  }{
    2\pi \hbar v_F^2 \rho \alpha
  }
  {\cal G}(\frac{q_c}{q_{j,T}}, \frac{k_F}{q_{j,T}})
  \\
  {\cal G}(a,b)
  &\equiv
  \int_{0}^\infty
  \frac{\dd{x}}{\epsilon_r(\frac{k_F x}{b})^2}
  \frac{
    \log\bigl(
    \frac{x}{a}
    +
    \frac{a}{x}
    \bigr)
    +
    1.3
  }{
    x^2 + 2a^2
  }
  \frac{
    x^3
  }{
    4\sinh^2(x/2)
  }
  \,.
\end{align}
Here, $q_{j,T}=k_BT/(\hbar{}s_j)$ are the thermal phonon wave vectors,
and $\Delta T=T_e-T_{\rm ac}$ is the temperature difference between electrons
and acoustic phonons.
The numerical factor in front of ${\cal G}$ is, taking
$D_2=\unit[7]{eV}$, $d_L\equiv{}c_LD_2^2/(2\pi\hbar v_F^2 \rho
\alpha)\approx0.008$ for longitudinal acoustic modes and
$d_T\approx0.003$ for transverse modes.  Moreover ${\cal G}$ diverges
logarithmically as $a\to0$ and $b\to0$, so that the total temperature
dependence is $T^2\ln^2(T)\Delta T$. In the opposite limit
$a\to\infty$ we have ${\cal G}(a,b)\sim{}a^{-2}$. For parameters of
graphene, the ratio is $q_c/q_{L,T}\sim{}\sqrt{\unit[100]{K}/T}$,
resulting to ${\cal G}$ of order 1 in the typical temperature
range as illustrated in Fig.~\ref{fig:Gfunc}.

Mechanical strain in graphene also cuts off the $q^2$ behavior of the
flexural phonon spectrum at low wave vectors.  For isotropic relative
strain $\tilde{u}$, one can find the corresponding result by replacing
$q_c\mapsto{}q_*=\sqrt{\tilde{u}}s_L/\alpha$, which is of the order of
$q_c$ for strains $\tilde{u}\sim{}10^{-4}$. Strain suppresses the
flexural phonon mediated energy transport in a similar way as it
suppresses the contribution to resistance.  \cite{castro2010-loc}

For the acoustic phonon process $w^{L,T}$, we can find the
supercollision energy transfer rate in a similar way:
\begin{align}
  {\cal J}
  &=
  2\pi N
  \frac{  
    \nu_1^2
    D_1^2D_2^2
  }{
    2 v_F^2 \rho^2 s_L s_T
  }
  \sum_{\gamma\gamma'}
  \int_0^\infty\frac{\dd{q}}{2\pi}
  \frac{
    q N^{el}_{\gamma\omega_{L,\vec{q}} + \gamma'\omega_{T,\vec{q}}}
  }{\epsilon_r(q)^2}
  \\\notag
  &\times
  (N_{\omega_{L,\vec{q}}} + \delta_{\gamma,+})
  (N_{\omega_{T,\vec{q}}} + \delta_{\gamma',+})
  (\gamma\omega_{L,\vec{q}} + \gamma'\omega_{T,\vec{q}})^2
  \,.
\end{align}
Neglecting screening and taking $s_T=0.67 s_L$, we have
\begin{align}
  \label{eq:ac-ac}
  {\cal J}_{LT}
  &\approx
  23
  N
  \frac{  
    g^2
    \nu_1^2
    k_B^3T^2
    \Delta T
  }{
    \hbar
  }
  \frac{
    D_2^2
    k_B T
  }{
    \hbar^2 v_F^2 \rho s_L^2
  }
  \,,
\end{align}
Screening reduces the numerical prefactor roughly by a factor of $(1 +
4k_F/q_{L,T})^{-2}$.

Finally, for the processes involving optical phonons, a
straightforward calculation for the transferred power density gives a
result small compared to the direct 1-phonon process
\cite{bistritzer2009-eci}.

\section{Discussion}

\begin{figure}
  \includegraphics{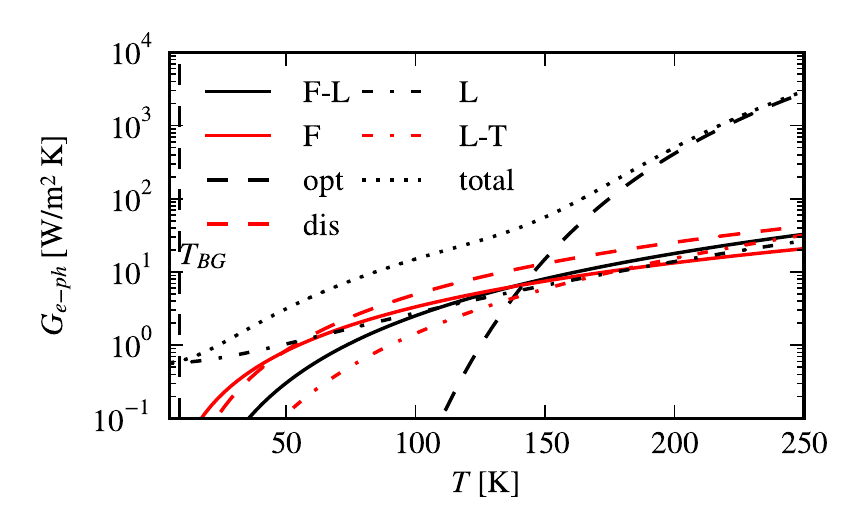}
  \caption{
    \label{fig:GJ}
    Thermal conductivity $G_{e-ph}={\cal J}/\Delta T$ between
    electrons and phonons, for monolayer graphene at charge density
    $n=\unit[10^{11}]{cm^{-2}}$, for parameter values given
    in the text. Several processes are shown:
    flexural-acoustic (F-L) multiphonon [Eq.~\eqref{eq:flex-ac}],
    acoustic-acoustic (L-T) [Eq.~\eqref{eq:ac-ac}],
    direct flexural phonon (F) [Ref.~\onlinecite{song2012-dae}],
    optical phonons (opt) [Ref.~\onlinecite{bistritzer2009-eci}],
    disorder-assisted supercollisions (dis) [Ref.~\onlinecite{song2012-dae},
    with $k_F\ell=200$], and acoustic phonons (L)
    [Ref.~\onlinecite{kubakaddi2009-imd,viljas2010-eph}].  Screening of deformation
    potential ($D_1=\unit[30]{eV}$) is in all processes taken into
    account as described in the text; this reduces the magnitude of
    the direct acoustic and flexural phonon processes, and slightly
    suppresses the others.
  }
\end{figure}

Similarly as for the resistance, the large population of the
low-wavevector flexural phonons plays an important role for the
flexural-acoustic phonon supercollisions. This is in contrast to the
direct process, \cite{song2012-dae} in which the most relevant
contribution comes from the thermal phonons with large wavevectors
$q_{F,T}\gg{}q_c,q_*$.  Although scattering from low-wavevector
flexural phonons occurs at a rapid rate, each event only transfers
energy $\hbar\omega_{F,\vec{q}_1}+\hbar\omega_{F,\vec{q}_2}$. In
contrast, a supercollision scattering event involving phonons of
similar wave vectors transfers a significantly larger energy
$\hbar\omega_{L,\vec{q}}$. This partly compensates for the smaller
matrix elements.

The energy flow for the direct flexural phonon process is \cite{song2012-dae}
\begin{align}
  \label{eq:flex-direct}
  {\cal J}_{\rm flex}
  =
  0.12 N \frac{[D_1(2k_F)^2+D_2^2] \nu_1(\mu)^2}{\rho^2\alpha^3}
  k_B^3T^2\Delta T
  \,,
\end{align}
where the screening of the deformation potential is taken into account
within the same model as above.  Comparing this to the multiphonon
process discussed above, with the parameter values discussed in
Sec.~II, we find ${\cal J}_L\sim{}{\cal J}_{\rm flex}$.  If screening
is neglected, we have instead ${\cal J}_L\sim0.1{\cal J}_{\rm flex}$.
The comparison to the direct flexural and the disorder-assisted
process \footnote{ Taking screening into account in disorder-assisted
  supercollisions \cite{song2012-dae} yields an effective
  electron-phonon coupling $\tilde{D}^2=D_1^2 F(k_F/q_{L,T}) +
  \frac{1}{2}D_2^2$ where $F(b)=\int_0^\infty\frac{\dd{x}}{2\zeta(3)
    \epsilon_r(k_Fx/b)^2}\frac{x^2}{e^x - 1}$. In the absence of
  screening, $F=1$. Moreover, note that the contribution of long-range
  Coulomb scattering to supercollisions is expected to be
  suppressed. \cite{song2011-sab} } is shown in Fig.~\ref{fig:GJ}. The
multiphonon process can be of a similar order of magnitude as the
disorder-assisted one in very clean samples.

Several parameters in Eq.~\eqref{eq:flex-ac} can be obtained by a
comparison to a resistivity measurement, in which flexural phonons in
the parameter regime relevant here contribute a $T^2$
increase. \cite{castro2010-loc,mariani2010-tdr} In particular,
assuming weak strain, $k_F,q_*\ll{}q_{F,T}$, one can rewrite
Eq.~\eqref{eq:flex-ac} in terms of the resistivity contribution
$\rho_F\propto{}T^2$ expected to originate from flexural phonons:
\cite{castro2010-loc,mariani2010-tdr,ochoa2011-tdr}
\begin{align}
  \label{eq:resistance-tau}
  {\cal J}_L
  &\approx
  100
  \frac{g^2 \nu_1^2 k_B^3 T^2 \Delta T}{\hbar}
  \frac{
    4e^2 \rho_F
  }{
    h
  }
  \begin{cases}
    \frac{q_*^2{\cal G}}{q_{L,T}^2}
    \,,
    &
    \text{$k_F\ll{}q_*$,}
    \\
    \frac{k_F^2{\cal G}}{q_{L,T}^2\ln\frac{q_{F,T}}{q_c}}
    \,,
    &
    \text{$q_*\ll{}k_F$,}
  \end{cases}    
\end{align}
where $g^2=D_1^2/(2\rho s_L^2)$. The electron-phonon coupling
$\tilde{D}$ in the resistivity in principle also contains the
deformation potential, but under the screening assumptions here and
using the parameter values quoted above (which are those used in
Ref.~\onlinecite{castro2010-loc}), we approximated
$\tilde{D}\approx{}D_2$.  What can be seen in
Eq.~\eqref{eq:resistance-tau} is that an effective $k_F\ell$ inferred
from the resistance for flexural phonons must be adjusted for the
difference in the characteristic phonon wave vector scales:
$\max(k_F,q_*)$ for resistance and $q_{L,T}$ for supercollisions.  For
the acoustic phonon process, we can find a relation to the resistance
contribution in a similar way: $(D_2^2k_B T)/(\hbar^2 v_F^2 \rho
s_L^2)\approx(4e^2 \rho_{L/T})/h$
(cf. e.g. Ref.~\onlinecite{ochoa2011-tdr}).

Flexural phonons are thermally induced dynamic ripples in graphene,
and as far as quasielastic scattering is concerned, the physics is
similar to the case of static ripples. Indeed, the function ${\cal
  Y}(\vec{q})$ in Eq.~\eqref{eq:Ycorr} above is closely related to the
correlation function of out-of-plane displacements that appears in the
case of static ripples. \cite{katsnelson2008-esm} Supercollision
scattering from static ripples was considered in
Ref.~\onlinecite{song2011-sab}, assuming small-scale ripples with
characteristic wave vector $q_c\gg{}q_{L,T},k_F$, and a
temperature-independent amplitude $Z$ chosen larger than the effective
thermal ripple size in the above flexural phonon
calculation. Interestingly, this also results to a $T^2\Delta T$
temperature dependence, but originating from the $a^{-2}$ scaling of
${\cal G}$.  The magnitude of the effect is sensitive to the amplitude
of the ripples: different experimental parameters
\cite{lundeberg2010-rgi,song2011-sab} lead to
$k_F\ell_{\rm{}eff}\sim{}a^2/q_c^2Z^4=10^0\ldots10^4$. The amplitude and
characteristic length scale of ripples is in principle visible also in
the resistivity. \cite{katsnelson2008-esm}

Finally, we can point out that electron--phonon processes in bilayer
graphene have similar phase space restrictions as in monolayer.
Supercollision and multiphoton processes are possible also there, and
have the advantage that due to the quadratic Hamiltonian, the two
terms in Eq.~\eqref{eq:rate} have the same sign, so that the selection
rules are less strict.  However, the magnitude of the effect is also
reduced, because the quadratic spectrum implies that the virtual
electron state involved lies at a higher energy. This gives more
weight to small $q$-values, which implies that screening will be of
importance.

In summary, I discussed the effect of multiphonon processes on the
energy transport in monolayer graphene at intermediate temperatures.
I find that in ultraclean suspended graphene samples, multiphonon
processes arising in second order of perturbation theory can compete
in the energy flow over first-order acoustic and flexural phonon
processes and disorder-assisted supercollisions. However, these
results are sensitive to the electronic screening, mechanical strain,
and non-thermal ripples in the system.

I thank P. Hakonen and T.T. Heikkil\"a for useful discussions.  This
work was supported by the Academy of Finland.

\end{document}